\documentclass[12pt]{article}

\usepackage{graphics,epsfig}

\def\dalemb#1#2{{\vbox{\hrule height .#2pt
        \hbox{\vrule width.#2pt height#1pt \kern#1pt
                \vrule width.#2pt}
        \hrule height.#2pt}}}

\let\a=\alpha   \let\d=\delta \let\e=\epsilon
\let\z=\zeta  \let\th=\theta  
\let\l=\lambda \let\m=\mu \let\n=\nu \let\x=\xi  

      \let\G=\Gamma  \let\Th=\Theta 
\let\X=\Xi  \let\S=\Sigma  \let\Y=\Psi
 
\let\la=\label  
  
\def\nn{\nonumber} \def\bd{\begin{document}} \def\ed{\end{document}}
\def\ds{\documentstyle} \let\fr=\frac \let\bl=\bigl \let\br=\bigr
\let\Br=\Bigr \let\Bl=\Bigl
\let\bm=\bibitem
\let\na=\nabla
\def\tU{{\widetilde U}}
\let\pa=\partial \let\ov=\overline
\def\ie{{\it i.e.\ }}
\newcommand{\be}{\begin{equation}}
\newcommand{\ee}{\end{equation}}
\def\ba{\begin{array}}
\def\ea{\end{array}}
\def\ft#1#2{{\textstyle{{\scriptstyle #1}\over {\scriptstyle #2}}}}
\def\fft#1#2{{#1 \over #2}}
\def\F#1#2{{ F_{#1}^{(#2)} }}
\def\cF#1#2{{ {\cal F}_{#1}^{(#2)} }}

\def\R{{\bf R}}
\def\sst#1{{\scriptscriptstyle #1}}
\def\oneone{\rlap 1\mkern4mu{\rm l}}
\def\e7{E_{7(+7)}}
\def\td{\tilde}
\def\wtd{\widetilde}
\def\im{{\rm i}}
\def\bog{Bogomol'nyi\ }
\newcommand{\ho}[1]{$\, ^{#1}$}
\newcommand{\hoch}[1]{$\, ^{#1}$}
\newcommand{\bea}{\begin{eqnarray}}
\newcommand{\eea}{\end{eqnarray}}
\newcommand{\ra}{\rightarrow}
\newcommand{\lra}{\longrightarrow}
\newcommand{\Lra}{\Leftrightarrow}
\newcommand{\ap}{\alpha^\prime}
\newcommand{\bp}{\tilde \beta^\prime}
\newcommand{\cB}{{\cal B}}
\newcommand{\cO}{{\cal O}}
\newcommand{\vecx}{\vec{x}}
\newcommand{\vecy}{\vec{y}}
\newcommand{\vecp}{\vec{p}}
\newcommand{\vecq}{\vec{q}}
\newcommand{\tr}{{\rm tr} }
\newcommand{\Tr}{{\rm Tr} }
\newcommand{\NP}{Nucl. Phys. }

\newcommand{\cL}{{\cal L}}
\newcommand{\cA}{{\cal A}}
\newcommand{\cD}{{\cal D}}

\def\sst#1{{\scriptscriptstyle #1}}
\def\0{{\sst{(0)}}}
\def\1{{\sst{(1)}}}
\def\2{{\sst{(2)}}}
\def\3{{\sst{(3)}}}
\def\4{{\sst{(4)}}}
\def\5{{\sst{(5)}}}
\def\6{{\sst{(6)}}}
\def\7{{\sst{(7)}}}
\def\8{{\sst{(8)}}}
\def\ve{\varepsilon}
\def\vf{\varphi}
\def\F{\Phi}
\def\wg{\wedge}

\newcommand{\tamphys}{\it 
}

\newcommand{\auth}{AUTHORS}

\def\thb{\bar{\theta}}
\def\Thb{\bar{\Theta}}
\def\barp{\bar{p}}
\def\barq{\bar{q}}
\def\barc{\bar{c}}
\def\bard{\bar{d}}
\def\e{\epsilon}

\def \bi{\bibitem}
\def \la {\label}

\def \l {\lambda}
\def\foot{\footnote}
\def \tl  {{\tilde \l}}
\def \sql {{\sqrt \l}}
\def \adss {$AdS_5 \times S^5$\ }
\newcommand{\rf}[1]{(\ref{#1})}
\def \ov {\over}

\def\th{\theta}
\def\Th{\Theta}
\def\vth{\vartheta}
\def\btheta{{\bar\theta}}
\def\ttheta{{{\tilde\theta}}}
\def\bttheta{{{\bar\ttheta}}}
\def\vth{\vartheta}

\def\ra{\rightarrow}
\def\N{{\cal N}}
\def\F{{\cal F}}
\def\uM{\underline{M}}
\def\uN{\underline{N}}
\def\uP{\underline{P}}
\def\cc{\circ}
\def\eqv{\equiv}

\def\ni{\noindent}

\def\Ep{E^{{}^{(+)}}}
\def\Em{E^{{}^{(-)}}}

\def\Mp{M^{{}^{(+)}}}
\def\Mm{M^{{}^{(-)}}}

\def \ha{{1\ov 2}}

\def\r{\rho}

\def\Y{{\rm Y}}
\def\X{{\rm X}}
\def\tY{\tilde{\rm Y}}
\def\tX{\tilde{\rm X}}
\def\dY{\dot{\rm Y}}
\def\dX{\dot{\rm X}}

\def \J {\mathcal{J}}
\def \del {\partial}

\def\dF{\dot{F}}
\def\dG{\dot{G}}
\def\df{\dot{f}}
\def \E {{\cal E}}
\def \S {{\cal S}}
\def \J {{\cal J}}

\def\ms{\mathcal{S}}
\def\mj{\mathcal{J}}
\def\soj{\fr{\ms}{\mj}}
\def \R {{\bf R}}
\def \om {\omega}
\def \bE {\bar E}
\def \x {{\cal X}}

\def \bi{\bibitem}
\def \la {\label}

\def \l {\lambda}
\def\foot{\footnote}
\def \tl  {{\tilde \l}}
\def \sql {{\sqrt \l}}
\def \adss {$AdS_5 \times S^5$\ }
\def \ov {\over}

\def \varpi {{\rm w}}

\def\thb{\bar{\theta}}
\def\Thb{\bar{\Theta}}
\def\zb{\bar{z}}
\def\psib{\bar{\psi}}
\def\barp{\bar{p}}
\def\barq{\bar{q}}
\def\barc{\bar{c}}
\def\bard{\bar{d}}
\def\e{\epsilon}

\def\At{\tilde{A}}
\def\Bt{\tilde{B}}
\def\ola{\overleftarrow}
\def\ora{\overrightarrow}
\def\at{\tilde{\a}}

\def\ps{\rlap{\, /}\;\,p }
\def\ks{\rlap{\, /}\;\,k }

\def\gym{g_{YM}}

\def\adot{\dot{a}}
\def\bdot{\dot{b}}

\newcommand{\PL}{{\em Phys.\ Lett.\ }}
\newcommand{\PR}{{\em Phys.\ Rev.\ }}
\newcommand{\PRP}{{\em Phys.\ Rep.\ }}
\newcommand{\CMP}{{\em Comm.\ Math.\ Phys.\ }}
\newcommand{\MPL}{{\em Mod.\ Phys.\ Lett.\ }}
\newcommand{\PRL}{{\em Phys.\ Rev.\ Lett.\ }}
\newcommand{\IJMP}{{\em Int.\ J.\ Mod.\ Phys.\ }}

\begin{document}
\overfullrule=0pt
\parskip=2pt
\parindent=12pt
\headheight=0in \headsep=0in \topmargin=0in
\oddsidemargin=0in

\vspace{ -3cm}
\thispagestyle{empty}

\begin{center}

{\Large\bf One loop scattering on D-branes
  }

 \vspace{.5cm} { I.Y. Park  }\\
 \vskip 0.2cm

{\it
Department of Chemistry and Physics, Philander Smith College\\
One Trudie Kibbe Reed Drive\\
Little Rock, AR 72202, USA \\
inyongpark05@gmail.com\\
}


\end{center}

 \vspace{0.1cm}

 \begin{abstract}
\ni  We analyze one loop scattering amplitudes of the massless
 states on a stack of D3-branes. We use the vertex operators that
 have been obtained in the direct open string analysis developed in
 arXiv:0708.3452. The method does not have the obstacle of the D9
 computation which is associated with the appearance of an $\e$-tensor.
 The divergence structure is not the same as the D9 brane case. What
 makes the analysis deviate from the D9 brane case is that the momenta
 of the states have non-zero components only along the brane directions.
 We ponder on the possibility that the one-loop divergence may be canceled
 by adding additional vertex operators at the tree level. We anticipate
 that they will be ``exponentiated'' to the free string action, with the
 resulting action to constitute a non-linear sigma model of the
 D-brane/AdS geometry.

\end{abstract}
\newpage

\setcounter{equation}{0}
\setcounter{footnote}{0}
\setcounter{section}{0}


\section{Introduction}
An open string is an interesting object for its end points among
other things. They are the places where the gauge symmetry  is
carried through the Chan-Paton factors. Potentially they can stick
together thereby converting the original open string into a closed
string. More recently it has been discovered that they may be
attached on a hyper-plane, a D-brane
\cite{pol,Polchinski:1995mt,Johnson:2000ch}. With its end points
attached, the open string will move on the D brane and may scatter
when it comes across another open string. Because of this it has to
be a due course of study to analyze various scattering amplitudes on
a stack of Dp branes, especially with $p<9$.\footnote{Scattering on
D-branes was studied in NSR formulation by several authors
\cite{Hashimoto:1996bf,bachas,Barbon,Lifschytz:1996a,gm,Balasubramanian:1996}.
We will use the Green-Schwarz formulation for a reason that will
become clear later. Also our approach is different in that we do not
introduce independent closed string fields.  } Certain pieces of
information on the amplitudes may be obtained by applying T-duality
to the results for the D9 brane. However, the range of the
information obtained in that manner is limited: many additional
pieces could be obtained by directly considering an open string on
lower dimensional D-branes. Furthermore a whole new picture seems to
emerge, as we propose below, on how the geometry would
arise\footnote{We emphasize that the geometry would arise {\em as a
result of the flat space computations}. More comments can be found
in sec 3. } as a way to cope with the open string loop divergences,
which does {\em not} have an analogue in the D9 brane physics. The
picture has implications to the open-closed string type dualities,
the matrix theory conjectures and AdS/CFT conjecture. Eventually it
will provide a new scheme for
unification of gauge theory and gravity.\\

The scattering physics on D-branes will obviously be relevant for
AdS/CFT correspondence and its generalization, which is in fact the
main motivation of the work. Although the methodology of the present
paper is general we will consider the D3 brane case to be specific.
In the stronger version of the AdS/CFT conjecture it is stated that
the $D=4\;\; \N=4$ SYM theory is fully equivalent,
without\footnote{This would really be the strongest version.} taking
any limit such as the large N limit, to the closed string theory on
AdS$_5\times S^5$. A low energy limit of an open string is $D=4\;\;
\N=4$ SYM: a SYM theory result should be acquired by taking a small
$\a'$ limit of the corresponding open string computation. In light
of the conjecture what it means is that the massive open string
modes do not play a role in producing the same results as those of
the dual closed string theory. This should be so in spite of the
fact that they are a natural (i.e., stringy) extension of the SYM.
During the past few years evidence along this line has been
collected. However the conjecture still remains a conjecture.
Furthermore there have been attempts to deduce or derive the
conjecture which only led toward its weaker form, but not
necessarily toward the stronger form
\cite{Park:1999xz,Kawai:2007ek}. (See \cite{DiVecchia:2005vm} also.)
Therefore we believe it is of prime importance to understand whether
(and if so, how) the full open stringy analysis figures into the
picture.\\

For that purpose it may be useful to consider scattering of open
strings on a stack of D3-branes. The first task will be construction
of vertex operators on D3-branes. In \cite{Park:2007mc} massless
vertex operators have been constructed and their tree level
scattering has been analyzed. One of the reasons why such an
independent analysis is necessary rather than relying on applying
T-duality on the D9 physics, is the non-commutativity of the quantum
corrections and T duality (or dimensional reduction) in the current
setting. An analogy may be helpful. In a standard quantum field
theory it is well-understood that there is no connection between the
quantum corrections of a dimensionally reduced theory and those of
the original theory. The reason is that the reduced theory loses
some degrees of freedom. One has a similar situation here. One can
easily see it from the world-sheet perspective. When computing the
one loop correction one takes the trace over the momentum, $\int
d^Dp<p|(\cdots)|p>$. In the D9 case one takes $D=10$. However for
the D3 case one should take $D=4$.\footnote{ Some cautionary remarks
may be needed here. In the conventional setup where one starts both
with an open string and a closed string, T-duality will commute with
the quantum corrections. This is because one would account for the
momentum and winding states and computes the quantum loops before
taking a large/small radius limit: the D9 brane results will
translate into the D3 brane results via T-duality. The current setup
is as if one takes the limit first and therefore the D3-brane
results cannot be recovered from the corresponding D9 results. Once
more an analogy with a quantum field theory may be useful. If one
compactifies on a circle and keeps all the momentum and winding
modes, there will be a connection between the two theories. However,
then one is not dealing with the dimensionally reduced theory. The
current setup of a pure open string is analogous to a dimensionally
reduced theory. For one thing we just saw that some of the zero
modes get lost. We believe that the correct approach to obtain the
D3 brane quantum effects is as we present in this paper, at least in
the purely open string setup. Whether such a purely open string
frame-work really exists is debatable. If the conjecture of this
paper can be verified to high loop orders, it will be an indication
that the answer is affirmative. }'\footnote{After this work was
published, it has been verified at the one-loop \cite{progress}. The
two loop extension has also been initiated more recently
\cite{Park:2009ki}.} This makes a difference in the divergence
behavior of the one loop. In the D9 case renormalization of the
string tension is sufficient to absorb the divergence. As we will
see below additional counter terms (presumably infinitely many of
them for all order cancellation) may be required in the
case of D3-branes. \\

The origin of the difference between the D9- and D3- analyses is
that the transverse momentum components are zero for the D3-brane
case \cite{Park:2007mc}.
 We extend the study to one loop and in particular consider
four point scattering amplitudes as shown in the figure below. The
outer boundary of the annulus is attached to the D3 branes as a
result of the Dirichlet boundary condition of an open string. The
inner boundary lies in the bulk. Since the momentum of the all four
open string states are along the longitudinal direction
\cite{Park:2007mc} it is natural to believe that inner boundary
should represent the closed string ``propagating" into the
transverse space {\it but with zero momentum}. Put another way the
closed string should be {\it non-propagating}. In this setting,
therefore, the status of a closed string is very different from that
of an open string. In terms of the low energy field theory it would
mean that the closed string would appear as an insertion of certain
composite operators, whereas an open string would be propagating,
fundamental degrees of freedom. Since an insertion of a closed
string vertex operator is associated, according to the common lore,
with change in the metric it is likely that the effect of the loop
is to deform it. To what metric will it deform? The only natural
candidate is the supergravity metric solution for a stack of D3
branes (or AdS$^5\times$S$^5$ when
the number of the branes is large).\\

\begin{figure}[!ht]
\centerline{
        \begin{minipage}[b]{7cm}
               \epsfxsize=5
               cm
                \epsfbox{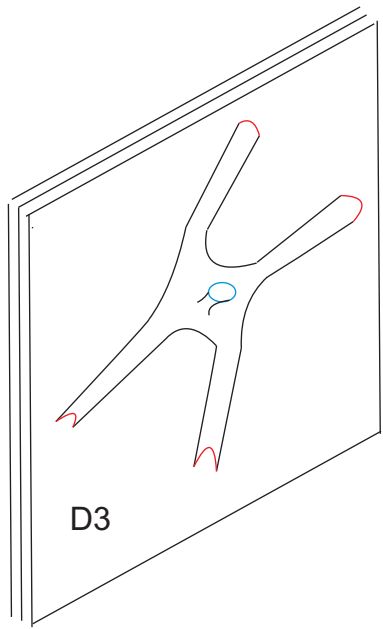}
        \end{minipage}
} \caption{ One loop four point open string scattering on a stack of
D3 branes} \label{loopfig}
\end{figure}

The amplitude computations will be performed in an operator
formulation. However, it is not the same as the one in \cite{gsw}:
instead of treating the first state and the last state as a
bra-state and a ket-state we put all vertex operators on an equal
footing.\footnote{ The motivation for considering such a formulation
is to have a convenient set-up for a future-check of the conjecture
that is put forward in sec 3. With the setup one can use the
standard Wick contraction techniques on the fields. The check will
require many lengthy amplitude computation with many vertex
operators inserted. In the existing operator formulation one must
deal with multiple oscillator products and it will be hard to
maintain a reasonable level of confidence in the accuracy of the
computations. It is absolutely advantageous to employ the standard
quantum field theoretic technique, the Wick contraction. It may not
be just a matter of convenience: as will be discussed in the
beginning of sec 2.}
 We insert the vertex operators constructed in \cite{Park:2007mc}. What
we find as a pleasant surprise is that
 the present formulation seems free of the well-known limitation
 of the light-cone gauge for D9 brane that is associated with
 setting $k^\pm=0$ for simplicity. With $k^\pm=0$ one cannot
 compute M-point amplitudes with $M>6$. The reason is the appearance
 of the $\e$-tensor with eight indices, which is not clear how to
 covariantize.\footnote{Presumably it is this $\e$ problem that the
 type of the formulation of this work was not used in the past.}
  As a matter of fact one should prove that such terms
 are absent as a separate task.  What saves the case of D3 brane (or for that matter
 other cases of Dp-branes with a low enough p) from similar
 difficulty is that the momenta have only two non-zero
components therefore making the $\e$-tensor term vanish. By treating
all the operators on an equal footing, the method has a more direct
link to the path integral approach.\footnote{The path integral that
we are referring to is not that of \cite{gsw} but will be a hybrid
approach of \cite{gsw} and \cite{pol}. The path integral approach of
\cite{gsw} is rather unwieldy in the sense that they rely on the
oscillator wave functions. The wave functions, especially the
fermionic ones, seems to be complicated.
 Here we use the ``conventional" vertex operators, i.e., the ones
 constructed \cite{Park:2007mc}.  They are the vertex operators in the
 Green-Schwarz formulation for the states on D3 branes. }
\\

 The rest of the paper is organized as follows: We start in Sec2 with a brief review
 of necessary ingredients. For a review of the string theory in general see
 \cite{gsw,kaku,pol}. To check the validity of the operator formulation
 with all operators on equal footing we
 compute various three- and four- point amplitudes at the tree level.
 Throughout the computations we use
dimensional regularization.
 In Sec3 we consider one loop four point amplitudes and work out the divergence
 structure. The
analysis goes differently from that of the D9 brane: it
 is not connected to the D9 result via T-duality. This is
  due to the fact that the momenta of the states are only along the D-branes
 but not along the transverse directions.
Motivated by a physical picture we propose a mechanism of divergence
cancellation by inserting additional vertex operators. We make a
schematic analysis for a quartic vertex operator for an illustration
that it produces terms with the correct pole structure. A more
complete study with the inclusion of a complete set of the vertex
operators will be given elsewhere. We also discuss how the geometry
might arise in this setting. We put forward a coherent way to view
AdS/CFT and similar type dualities assuming the validity of the
proposed picture. We conclude with discussions of a few other issues
and future directions.

\section{Computations with operators on equal footing }
In this section we discuss the computations of various amplitudes in
a setting where all the vertex operators are treated on an equal
footing. In other words all of them are inserted between a bra- and
 ket- states which are the {\em vacuum} states. The set-up will
provide a convenient stage for a future-check of the conjecture that
we put forward in section 3.

 An M-point amplitude in general is given by
 \bea
 A_M=\int d\m
 <\prod_{i=1}^{M}\;V(k_i)>
 \eea
The measure $d\m$ is given by
 \bea
 d\mu=|(x_1-x_2)(x_{1}-x_{M})(x_2-x_{M})|
 \int dx_3...dx_{M-1}\prod_1^{M-1}\th(x_{r
 }-x_{r+1}) \label{dmu}
 \eea
\ni The vertex operators for the massless states have been obtained
in \cite{Park:2007mc}: we refer to \cite{Park:2007mc} for them and
for our conventions.
The bosonic and the fermionic propagators are respectively
 \bea
 <X^i X^j>&=& -2\a'\eta^{ij} \ln|x-x'| \nn\\
 <S_1^{a_1}S_1^{a_2}>&=& \fr{\d^{a_1a_2}}{x_1-x_2}
 \label{ss}
 \eea

\subsection{four point amplitudes}

One can apply the formulation to various three point amplitudes. We
do not present the result, instead refer the interested readers to
\cite{Park:2008sg}. We turn to the various four point amplitudes.
Recall the limitation of setting $k^\pm=0$ in the operator
formulation of a D9 brane \cite{gsw,kaku}. It does not allow one to
compute M-point amplitude with $M>6$: with $M>6$ one encounters an
$\e$-tensor with 8 indices and it is not clear how to covariantize
the result. A pleasant surprise is that the limitation is absent in
the current formulation of a D3 brane where the $\e$-tensor appears
in the four point amplitude already. Basically the reason is that
the present approach treats the first state and the final state on
equal footing with all the other states, whereas in the operator
method they appear as a bra and ket respectively. Therefore one
expects to face the $\e^{(8)}$ issue already at a four point level.
What saves the formulation is that with the D3-brane the $\e$-term
vanishes due to the fact that the momenta only have two non-zero
components in the brane directions. Through several examples below,
we again will demonstrate that the present formulation yields the
same results as the conventional operator method.
\\

\ni Let's consider four point amplitudes in the order of increasing
complexity. The simplest is the four scalar amplitude.

 \bea
 I_{\phi\phi\phi\phi}=
 <\prod_{i=1}^4\; (\xi^{m_i}{X'}^{m_i}
 +\xi^{m_i}R^{m_iv_i}k^{v_i})
 e^{ik_i\cdot X}>
 \eea
There are six different types of terms: $XXXX, XXXR, XXRR,XRRR\;
\mbox{and}\; RRRR.$ In the dimensional regularization the second
type of terms vanish.
 The result is precisely the same as the result of the operator
formulation \cite{Park:2007mc}. Omitting the factor
$\fr{g^2}{2}\a'^2\;\Tr(\l^a\l^b\l^c\l^d)
\fr{\G(-s/2)\G(-t/2)}{\G(1-s/2-t/2)}$ one gets
 \bea
 A_{\phi\phi\phi\phi}=\fr14
 \left(su\;\xi_1\cdot \xi_4\;\xi_2\cdot \xi_3
 +tu\;\xi_1\cdot \xi_2\;\xi_3\cdot \xi_4
 +st\;\xi_2\cdot \xi_4\;\xi_1\cdot \xi_3
 \right)
 \eea
Many terms that would be otherwise present vanish due to the fact
that $\xi\cdot k=0 $. It is more involved to compute the two vector
and two scalar scattering amplitude. We illustrate this with
 \bea\label{faaf}
 I_{\phi AA\phi}=
 <\prod_{i=1,4}\;(\xi^{m_i}{X'}^{m_i}
 +\xi^{m_i}R^{m_iv_i}k^{v_i})e^{ik_i\cdot X}\prod_{i=2}^3\; (\z^{u_i}\dot{X}^{u_i}
 -\z^{u_i}R^{u_iv_i}k^{v_i})e^{ik_i\cdot X}>
 \eea
The corresponding operator result in the conventional formulation,
apart from $\fr{g^2}{2}\tr( \l^a\l^b\l^c\l^d)$, is
\cite{Park:2007mc}
 \bea
&& <k^1,\xi^1|V_g(k^2,\z^2)V_g(k^3,\z^3)|k^4,\xi^4 >\nn\\
=&& \xi^1\xi^4\left[\fr14\; su\;\z^2\cdot\z^3-\fr12 ( u\;\z^2\cdot
k^1\;\z^3\cdot k^4 + s\;\z^2\cdot k^4\;\z^3\cdot k^1) \right]
 \label{op-ffff}
 \eea
 where we have omitted the factor $\fr{\G(-\a' s)\G(-\a'
 t)}{\G(1-\a's-\a't)}$.
We break the computation of (\ref{faaf}) into pieces. First we check
the coefficient of the $\xi^1\cdot \xi^4\;\z^2\cdot\z^3$ term, and
subsequently the remaining terms. In all of the following
computations we omit the common factor $\fr{\G(-\a' s)\G(-\a'
 t)}{\G(1-\a's-\a't)}$. There are three contributions:
 \bea
 XXXX & \Rightarrow &  \fr{su/4}{1+t/2} \nn\\
 XXRR & \Rightarrow &  {t}\fr{su/4}{1+t/2} \nn\\
 RRRR & \Rightarrow & \fr{t^2}{4}\fr{su/4}{1+t/2}-\fr18 stu
 \eea
 They add up to yield
 \[\fr14 su\]
 which is indeed the correct
coefficient. The results for the remaining terms can be summarized
similarly. Unlike above the $XRRR$-terms contribute. One can work
them out explicitly using the identities given in the appendix. The
result is
 \bea
 XXXX & \Rightarrow &
 -\fr12\left(u\,\z^2\cdot k^1\;\z^3\cdot k^4+
   s\,\z^2\cdot k^4\;\z^3\cdot k^1
   +\fr{su/2}{1+t/2}\;\z^2\cdot k^3\;\z^3\cdot k^2\right) \nn\\
 XXRR & \Rightarrow &   \fr{su/4}{1+t/2}\;\z^2\cdot k^3\;\z^3\cdot k^2
 -\fr{t}{2}\fr{su/4}{1+t/2}\;\z^2\cdot k^3\;\z^3\cdot k^2 \nn\\
               &&        -\fr14\left(t^2\,\z^2\cdot k^4\;\z^3\cdot k^4
             -tu\,\z^2\cdot k^3\;\z^3\cdot k^4
             -ts\,\z^2\cdot k^4\;\z^3\cdot k^2\right)\nn\\
 XRRR & \Rightarrow &  \left(\fr14 u^2\;\z_2\cdot k_3\;\z_3\cdot k_4
           -\;\fr14su\;\z^2\cdot k^3\;\z^3\cdot k^1
           -\;\fr14 tu\;\z^2\cdot k^4\;\z^3\cdot k^4 \right.\nn\\
          && \left.+\fr14ts\;\z^2\cdot k^4\;\z^3\cdot k^1
            +\fr14 s^2\;\z^2\cdot k^4\;\z^3\cdot k^2
             -\fr14 su\;\z^2\cdot k^1\;\z^3\cdot k^2 \right. \nn\\
           && \left.  -\fr14 ts\;\z^2\cdot k^4\;\z^3\cdot k^4
             +\fr14 tu\;\z^2\cdot k^1\;\z^3\cdot k^4 \right) \nn\\
 RRRR & \Rightarrow & \fr{t}{2} \fr{su/4}{1+t/2}\;\z^2\cdot k^3\
                 ;\z^3\cdot k^2-\fr12\;su\;\z^2\cdot k^3\;\z^3\cdot k^2
       \nn\\
       && -\fr14(st\;\z^2\cdot k^4\;\z^3\cdot k^1
       +tu\;\z^2\cdot k^1\;\z^3\cdot k^4)
 \eea
 where we have omitted the common factor $\xi^1\cdot \xi^4$. Adding
 all four contributions one reproduces the last two terms of
 (\ref{op-ffff}),
 \bea
  -\fr12\left(u\,\z^2\cdot k^1\;\z^3\cdot k^4+
   s\,\z^2\cdot k^4\;\z^3\cdot k^1 \right)
 \eea
Note that $XXXX,XXRR,XRRR,RRRR$ terms individually produce terms of
the type $\xi\cdot k\;\xi\cdot k\;\z\cdot k\;\z\cdot k$ which equal
zero since $\xi\cdot k=0$. This completes the discussion of the two
scalar and two vector amplitude. The computation of the four vector
amplitude
 \bea
 I_{4v}=  
 <\prod_{i=1}^4\; (\z^{u_i}\dot{X}^{u_i}
 -\z^{u_i}R^{u_iv_i}k^{v_i})\;
 e^{ik_i\cdot X}>
 \eea
goes parallel although it is more involved. There are four different
kinds of terms as before. The computation of $\z\cdot \z\;\z\cdot
\z$-type of terms is similar to those of above. Let's consider the
terms of the form $\z\cdot k\;\z\cdot k\;\z\cdot \z$. To be specific
we take the example of $\z_1\cdot \z_2$. The results can be
summarized as
 \bea
XXXX &\Rightarrow& -\a'\left(u\;\z_3\cdot k_2\;\z_4\cdot
k_1+t\;\z_3\cdot k_1\;\z_4\cdot
k_2\right)-\fr{\a't\a'u}{1+\a's}\;\z_3\cdot k_4\;\z_4\cdot k_3\nn\\
 XXRR&\Rightarrow& \fr{\a't\a'u}{1+\a's}\;\z_3\cdot k_4\;\z_4\cdot k_3\\
 && -\left(\fr14 s^2\;\z_3\cdot k_2\;\z_4\cdot k_2-\;\fr14
 su\;\z_3\cdot k_2\;\z_4\cdot k_3-\;\fr14 st\;\z_3\cdot
 k_4\;\z_4\cdot k_2\right.\nn\\
 &&\left.+ \fr{s}{2}\;\fr{\a't\a'u}{1+\a's}\;\z_3\cdot
 k_4\;\z_4\cdot k_3\right)\nn
 \eea
 \bea
 XRRR  &\Rightarrow&
  \left(
  -\fr14 st\;\z_3\cdot k_2\;\z_4\cdot k_2+\;\fr14 su\;\z_3\cdot
 k_2\;\z_4\cdot k_1+\;\fr14 t^2\;\z_3\cdot k_4\;\z_4\cdot k_2 -\fr14
 tu\;\z_3\cdot k_4\;\z_4\cdot k_1\right.\nn\\
&&\left. -\fr14 us\;\z_3\cdot k_2\;\z_4\cdot k_2+\fr14 ts\;\z_3\cdot
 k_1\;\z_4\cdot k_2+\fr14 u^2\;\z_3\cdot k_2\;\z_4\cdot k_3-\fr14
 tu\;\z_3\cdot k_1\;\z_4\cdot k_3 \right)\nn\\
  RRRR  &\Rightarrow&
  \fr{s}{2}\;\fr{\a't\a'u}{1+\a's}\;\z_3\cdot k_4\;\z_4\cdot k_3
-\fr12\;tu\;\z^3\cdot k^4\;\z^4\cdot k^3\nn\\
&&-\left(\fr14 su\;\z_3\cdot k_2\;\z_4\cdot k_1+\;\fr14
st\;\z_3\cdot k_1\;\z_4\cdot k_2\right)
 \eea
Adding all four contributions one gets as the coefficient of
$\z^1\cdot \z^2$
 \bea
-\fr12\left(u\;\z_3\cdot k_2\;\z_4\cdot k_1+t\;\z_3\cdot
k_1\;\z_4\cdot k_2\right)+\fr{1}{8}t u\;\z_3\cdot k_4\;\z_4\cdot k_3
 \eea
It is precisely the same result as the one obtained in the
conventional operator method. Finally one can show that the terms of
the type $\z\cdot k \;\z\cdot k\;\z\cdot k\;\z\cdot k$ completely
cancel among themselves, confirming the conventional operator result
that those types of terms are absent.

\section{Analysis of the one loop divergence  }

The above method should be applicable to one loop computations. We
will not pursue it here. Rather we use the conventional operator
method \cite{gsw,kaku} which is simpler, to find various one loop
amplitudes. The divergence structure is different from that of the
D9. As has been emphasized previously it is due to the fact that the
Fock space momentum takes the non-zero values only in the brane
directions. After obtaining the one loop amplitudes we ponder on how
to remove the divergence. We will carry out some preliminary check
to see whether it is possible to remove it by adding counter terms
in the action, such as quartic $X$-terms, and evaluating their
contributions at the tree level. For that we again will turn to the
present operator method. We make a conjecture on how the
counter-terms may be connected to the geometry.

\subsection{one loop divergence  }

\ni In the operator method of \cite{gsw,kaku}, the one loop
divergence can easily be computed in analogy with the D9 brane case.
The only difference occurs in the bosonic zero modes, i.e., the
momentum, when one takes the trace. The difference, although looking
minor, is crucial since it is what makes the factor of $(\ln w)$
below different from the D9 case, which in turn changes the pole
structure. The divergence cancellation mechanism through an
introduction of geometry (which we propose below) hinges on this
difference.\\

\ni In close analogy with the D9 brane computation one can show that
a one-loop four point amplitude of the massless states in general is
given by
 \bea
 A(1,2,3,4)=g^4GK\int\fr{dw}{w}\int\;\left(\prod_{r=1}^3
 \fr{d\r_r}{\r_r}\right)\left(\fr{-2\pi}{\ln w}\right)^2\prod_{r<s}
 (\psi_{rs})^{k_r\cdot k_s} \label{1L}
 \eea
 where $G$ is the group theory factor
 \bea
&&
\Tr(\l^a\l^b\l^c\l^d)+\Tr(\l^a\l^d\l^c\l^b)+\Tr(\l^a\l^c\l^b\l^d)\nn\\
+&&\Tr(\l^a\l^d\l^b\l^c)+\Tr(\l^a\l^b\l^d\l^c)+\Tr(\l^a\l^c\l^d\l^b)
 \label{v4ptleading}
 \eea
The factor $K$ is a kinematic factor and depends on the states under
consideration. For the four vector amplitude for example
 it is given by
 \bea
 K=t_{i_1j_1i_2j_2i_3j_3i_4j_4}\z_1^{i_1}\z_2^{i_2}\z_3^{i_3}\z_4^{i_4}
 k_1^{j_1}k_2^{j_2}k_3^{j_3}k_4^{j_4}
 \eea
 with
 \bea
 t_{i_1j_1i_2j_2i_3j_3i_4j_4}=
 \Tr(R_0^{i_1j_1}R_0^{i_2j_2}R_0^{i_3j_3}R_0^{i_4j_4})
 \eea
For the four scalar scattering it is
 \bea
 -\fr14
 \left(su\;\xi_1\cdot \xi_4\;\xi_2\cdot \xi_3
 +tu\;\xi_1\cdot \xi_2\;\xi_3\cdot \xi_4
 +st\;\xi_2\cdot \xi_4\;\xi_1\cdot \xi_3
 \right)\nn\\
 \eea
For other amplitudes the $K$ should be replaced appropriately.
 Note that the power of $\left(\fr{-2\pi}{\ln
 w}\right)$ is different from the D9: instead of five it is two
 for the D3 case.\footnote{ To compare with the D9, the divergence structure of
 (\ref{1L}) can be put in the following variables
 \bea
 q=e^{\fr{2\pi^2}{\ln w}}, \quad \n_r=\fr{\ln \r_r}{\ln w}
 \eea
Eq (\ref{1L}) now takes the form of
 \bea
 A(1,2,3,4)=g^4GK(-8\pi^4)\int\fr{dq}{q(\ln q)^3}\;F(q^2)\label{1Lq}
 \eea
where
 \bea
F(q^2)= \int\;\prod_{r=1}^3 {d\n_r} \prod_{1\leq r<s \leq 4}
\left[\sin \pi(\n_s-\n_r)\prod_{n=1}^\infty
 (1-2q^{2n}\cos 2\pi (\n_s-\n_r)+q^{4n})\right]^{k_r\cdot k_s}
 \eea
Compared with the D9 where the logarithmic factor disappears, the
degree of divergence is more serious as
 \bea
 \int\fr{dq}{q(\ln q)^3}
 \eea
 }
 Note that the divergence of (\ref{1L}) comes from $w\sim 1$ and has
 the pole structure of
 \bea
 \int_0^1 dw\;\fr{1}{w(\ln w)^2} \label{1Lpole}
 \eea
 In the next section we discuss a possible
mechanism to cancel the divergence after discussion of physical
motivation. Before we get to that we express, for comparison later,
the pole structure of (\ref{1Lpole}) in a new coordinate, $\ln
w\equiv -y $: the pole structure takes the form of
 \bea
 \int_0^\infty dy\;\fr{1}{y^2} \label{ypole}
 \eea

\subsection{ anticipated mechanism for divergence cancelation}
The presence of the $(\ln q)^3 $-factor in (\ref{1Lq}), which does
not have an analog in the D9 case, seems to suggest a more radical
measure for the divergence cancellation. We discuss the possibility
that the one-loop divergence may be cancelled with additional vertex
operators at the tree level. We expect that they originate from the
curved geometry in the sense explained below.

Let's go back to the figure in the introduction where we have
pointed out that the status of a closed string is different from
that of an open string. Whereas the open strings are the fundamental
degrees of freedom, the closed strings are not in a several regards:
first of all they have been {\em generated} by open string quantum
effects. Secondly, they seem to be non-propagating due to the
momentum restriction on the open string states. Since it is not
fundamental degrees of freedom a natural way to realize them should
be via composite\footnote{\ni The open string vertex operators
themselves take the forms of the composite operators. Therefore what
is meant by composite here is that they are even more composite
and/or of different compositeness than the open string ones.}
operators. The crucial question is then, how should they be
introduced? Here we propose that they be introduced in such a way to
cancel (at least potentially) the open string divergences, in other
words, as counter terms.

The connection to the geometry may occur when they get
``exponentiated" to the action where they would appear as vertex
terms.\footnote{It will be more clear to see in the path-integral
approach where those composite operators will be brought down by the
standard procedure.} Together with the quadratic part of the action
they will constitute a non-linear sigma model action. We emphasize
that the geometry would arise {\em as a result of the flat space
computations}. The counter vertex operators will be introduced in
attempt to remove the divergence that has resulted from the {\em
flat} space computation. (For example we are {\em not} quantizing
the open string in a curved space. As a matter of fact, if the
conjecture is true it will actually circumvent the necessity of
quantizing the string on a curved space, at least for the open
string.) It is similar, in spirit, to what was shown in
\cite{Gonzalez-Rey:1998uh,Kruczenski:2006ti}. There it was shown
that the effective action of N=4 SYM that contains quantum (and the
non-perturbative ) corrections can be interpreted as a non-linear
sigma type action with a curved target space. If this anticipation
turns out true to this point then we believe that the resulting
action should be an action for an open string in a curved geometry.
To what geometry would it deform? The only natural candidate is the
supergravity metric solution for a stack of D3 branes or
AdS$^5\times$S$^5$ in an appropriate limit. Stated the other way
around the precise set of the vertex operators will be dictated by
the non-linear sigma model action and the divergence cancellation.
Again we emphasize that the curved geometry should be used in order
to guide us in finding the forms of the vertex operators, but not as
a space to quantize an open string in. Finding the forms of the
counter vertex operators without such guidance would be a hopeless
task.

Although the counter terms may reveal aspects of the geometry which
part of the geometry is revealed may depend on the dynamics
considered. Another possible question is on the realization of the
closed string. In the current stage they reveal their presence
through the deformed metric. Could they appear on a more fundamental
level? We postpone these issues until the conclusion. In the
remainder of this section we carry out a preliminary check to set
the ideas above on a computational ground. For an illustration we
take the example of the four point scalar scattering.\footnote{The
corresponding calculation with the vector vertex operators would be
more involved. For this reason and others the advantage of having
the explicit forms of the vertex operators is obvious. }\\

\ni The one loop divergence structure has been presented in the
previous section. One needs to find the complete list of the vertex
operators that would cancel the divergence. The task \cite{progress}
involves lengthy computations which involve many vertex operators.
Here we only focus on one of them to illustrate the
strategy.\footnote{To cancel the divergence using the conventional
operator formulation along the line of the conjecture of the present
work, the only natural thing to do is to change the Virasora
operator $L_0$ to take the "curved space" effects into account. It
is not entirely clear, apart from the technical complexities,
whether there will be a valid series expansion in this approach. }
Recall that a curved metric has the following curvature expansion in
a Riemann normal coordinate,
 \bea
 g_{MN}=\eta_{MN}-\fr13 R_{MPNQ}(X_0)X^PX^Q+\cdots
 \eea
The supergravity metric solution for a stack of D3 branes is given
by
 \bea
 ds^2 &=& H^{-1/2}(-dt^2+(dx^\m)^2)+H^{1/2}(dx^m)^2\nn\\
  H &=& 1+\fr{4\pi gNl^4}{r^4}
 \eea
Consider a $r$-expansion of the curvature tensor, $R_{mnpq}$, that
results from the D3-brane metric where we treat $r$ to be a large
but fixed constant. In the leading order the term that has the least
number of fields is
 \bea
 R_{mnpq}\sim (\d_{mq}\d_{np}-\d_{mp}\d_{nq})
 \eea
 Therefore one of the vertex operators that is necessary to cancel
 the one loop divergence from $<V_s^{m_1}V_s^{m_2}V_s^{m_3}V_s^{m_4}>$
is expected to have the following form
 \bea
 \int dx\; C_{mpnq}\, \dot{X}^m \dot{X}^n X^p X^q
 \eea
When inserted together with the other vertex operators that
represent the scattering states, the tree level amplitude
 \bea
 \int d\m <V_s^{m_1}(x_1)V_s^{m_2}(x_2)V_s^{m_3}(x_3)V_s^{m_4}(x_4)
 \int dx\; C_{mpnq}\, \dot{X}^m(x) \dot{X}^n(x) X^p(x) X^q(x)>
 \label{ctr}
 \eea
produce terms that have the same pole structure as that of the one
loop. The measure in front $d\m$ is given in (\ref{dmu}) and
 \bea
  C_{mpnq}=\mbox{const} \cdot (\d_{mq}\d_{np}-\d_{mn}\d_{pq})
 \eea
The constant will be a function of the t' Hooft coupling and will be
determined by the requirement of the divergence
cancellation.\footnote{It will also contain a certain (positive or
negative) power of $r$. Presumably one should treat $r$, but not the
individual coordinates, as a constant as in the case of a point on a
sphere of radius $r$. This point of view already appeared in the
context of AdS/CFT \cite{Das:1998ei}. The detailed mechanism to
treat $r$ should be a part of how open string quantum effects reveal
geometry. (The detailed mechanism has now been given in
\cite{progress}.)} The amplitude (\ref{ctr}) have poles at
 \[
x=x_i,\quad i=1,..,4
 \]
The pole terms at $x=x_1$ vanishes as $x_1\ra \infty$. The highest
order pole terms at $x=x_i, i\neq 1$ have the structure of
 \bea
 \int_0^\infty dy\;\fr{1}{y^2} \label{ypoleprime}
 \eea
where we have not recorded the precise form of the coefficient. Note
that it is the same order pole as the one loop divergence given in
(\ref{ypole}). Eq.(\ref{ctr}) also produces terms of a different
pole structure, $\int_0^\infty dy\;\fr{1}{y}$, and/or terms that
become divergent as $x_1\ra \infty$. It is crucial to check that all
these unwanted terms must cancel among themselves when the complete
list of the vertex operators are once considered together. At the
same time the momenta structure must turn out to match that of the
one loop for the correct pole terms.  We have elaborated on this in
\cite{progress}.

\section{Conclusion}

In this letter we have analyzed the one loop divergence structure of
the four point scattering amplitudes. We have conjectured the
existence of a complete set of additional vertex operators whose
origin should be linked to the D-brane/AdS geometry. In weak
coupling the relevance of the closed strings seems to be recognized
rather indirectly, i.e., through the non-linear sigma model. But is
there a circumstance where they become propagating degrees of
freedom?  We believe that it is when one goes to a large coupling
limit where closed string degrees of freedom become fundamental. It
should occur through an open string conversion into a closed string
\cite{Park:2001bm}. Then it will suggest that AdS/CFT type dualities
will be a two-step process. First the open string quantum
corrections will ``engineer" the curved geometry, which can be
viewed as an open string generalization of the gauge theory result
of \cite{Chepelev:1997fk,Periwal:1998sn,Gonzalez-Rey:1998uh}. The
curved geometry will be introduced as a way to absorb the
divergence. At the same time it will serve as a route for the closed
strings to exhibit their relevance. At this stage the closed strings
are not fundamental degrees of freedom but they become so when one
reaches a strong coupling region through an S-duality, as an open
string converts to a closed string.\footnote{One may wonder about
the reverse mechanism where a closed string converts into an open
string. We suspect that it should be a process that involves a
spontaneous symmetry
breaking. This issue will be pursued elsewhere.  } \\

 A few comments on the future
directions are in order : One obvious direction  is to find those
set of the vertex operators, at least order by order. For that
purpose the works of \cite{Metsaev:1998it,Cvetic:1999zs} will
provide a useful guide. Another direction is to fully develop the
path integral formulation. The setting of the present work is such
that one can readily switch to the path integral method. There the
vertex ``operators" will appear in the standard procedure from the
exponentiated action. If our picture is indeed correct one can say
that the open string dynamics reveals aspects of the geometry. Which
particular part of the geometry becomes revealed depends on the
dynamics considered. For example, considering scalar multiple
scattering will reveal different pieces of information about the
geometry that the vector scattering. So the counter-terms necessary
to cancel the divergences will be the geometry information relevant
for the dynamics. Also there is a question concerning the radius of
the sphere that originates in the large N-limit of the open string
engineered geometry. Since there is an S-duality involved to go to
the strong coupling it may be the inverse (in terms of the t'Hooft
coupling) of the radius of the regular sphere that results from the
D3 brane supergravity solution. \\
\indent Connection between the string divergence cancellation and
the field theory cancellation may be an interesting issue as well.
The field theory task has been initiated in \cite{Park:2007ev} where
the starting point is the four dimensional action with
$\a'$-corrections. The action is obtained by dimensional reduction
of the ten dimensional action \cite{Gates:1986is}. Since the massive
modes have been integrated out the ``renormalization" process will
go differently from the full fledged string analysis. However we
expect that they might be properly taken into account by an energy
scale. We
hope to report on this with better understanding in the future.\\

\ni Finally a possible connection to the Fischler-Susskind
mechanism: A few ingredients (such as the role of the zero momentum
states) of the conjectured divergence cancelation are reminiscent of
Fischler-Susskind mechanism \cite{Fischler:1986ci,Das:1986dy}. One
of the differences is the setting: here we start out only with the
open string degrees of freedom. Existence of such a setup is not
fully established although it is one of the issues that is pursued
by the string field theory. According to the string theory common
lore, an open string always needs a closed string. However what we
are conjecturing is independent of the frameworks: we are
conjecturing that the open string divergence should be removed by
purely open string vertex operators. If it does not work this way,
then closed string vertex operators may be considered. In other
words, in the conventional set-up where one includes the closed
string degrees of freedom as well, one would try to cancel the
divergence by some closed string tadpole divergence. However, it
does not seem natural to attempt to remove the divergence of a
single integral in (\ref{ypole}) with a $dyd\bar{y}$-type double
integral that the closed string analysis will produce. Furthermore
there is another string theory common lore that states that an
oriented open string cannot be coupled to an oriented closed string,
although the statement is not being brought up too often with the
advent of D-brane physics.
\\



\begin{thebibliography}{20}

\bibitem{pol}
  J. Plochinski,
  String theory, vol 1,2, Cambridge

\bibitem{Polchinski:1995mt}
  J.~Polchinski,
  ``Dirichlet-Branes and Ramond-Ramond Charges,''
  Phys.\ Rev.\ Lett.\  {\bf 75}, 4724 (1995)
  [arXiv:hep-th/9510017].

\bibitem{Johnson:2000ch}
  C.~V.~Johnson,
  ``D-brane primer,''
  arXiv:hep-th/0007170.







\bibitem{Hashimoto:1996bf}
  A.~Hashimoto and I.~R.~Klebanov,
  ``Scattering of strings from D-branes,''
  Nucl.\ Phys.\ Proc.\ Suppl.\  {\bf 55B}, 118 (1997)
  [arXiv:hep-th/9611214].
 ;
I.\ R.\ Klebanov and L.\ Thorlacius, \PL {\bf B371} (1996) 51-56
 ;
S.S.\ Gubser, A.\ Hashimoto, I.R.\ Klebanov, and J.M.\ Maldacena,
  ``Gravitational lensing by $p$-branes,'' \NP {\bf 472} (1996) 231-248, {\tt
  hep-th/9601057}
 ;
A.\ Hashimoto and I.R.\ Klebanov, ``Decay of Excited D-branes,'' \PL
{\bf B381}
  (1996) 437-445, {\tt hep-th/9604065}
 ;
A.\ Hashimoto, ``Dynamics of Dirichlet-Neumann Open Strings on
D-branes,'' {\tt
  hep-th/9608127}
 ;
A.\ Hashimoto, ``Perturbative Dynamics of Fractional Strings on
Multiply Wound
  D-branes,'' {\tt hep-th/9610250}



\bibitem{bachas}
C.\ Bachas, ``D-Brane Dynamics,'' \PL {\bf B374} (1996) 37-42, {\tt
  hep-th/9511043}

\bibitem{Barbon}
J.\ Barbon, ``D-Brane Form-Factors at High-Energy,'' \PL {\bf B382}
(1996)
  60-64, {\tt hep-th/9601098}

\bibitem{Lifschytz:1996a}
G.\ Lifschytz, ``Comparing D-branes to Black-branes,'' {\tt
hep-th/9604156}

\bibitem{gm}
M.\ R.\ Garousi and R.\ C.\ Myers, ``Superstring Scattering from
D-Branes,''
  \NP {\bf 475} (1996) 193-224, {\tt hep-th/9603194}

\bibitem{Balasubramanian:1996}
V.\ Balasubramanian and I.\ R.\ Klebanov, ``Some Aspects of Massive
World-Brane
  Dynamics,'' {\tt hep-th/9605174}

















\bibitem{Park:1999xz}
  I.~Y.~Park,
  ``Fundamental vs. solitonic description of D3 branes,''
  Phys.\ Lett.\ B {\bf 468}, 213 (1999)
  [arXiv:hep-th/9907142].

\bibitem{Kawai:2007ek}
  H.~Kawai and T.~Suyama,
  ``AdS/CFT Correspondence as a Consequence of Scale Invariance,''
  arXiv:0706.1163 [hep-th].


\bibitem{DiVecchia:2005vm}
  P.~Di Vecchia, A.~Liccardo, R.~Marotta and F.~Pezzella,
  ``On the gauge / gravity correspondence and the open/closed string
  duality,''
  Int.\ J.\ Mod.\ Phys.\  A {\bf 20}, 4699 (2005)
  [arXiv:hep-th/0503156].



\bibitem{Park:2007mc}
  I.~Y.~Park,
  ``Scattering on D3-branes,''
  arXiv:0708.3452 [hep-th], to appear in  Phys. Lett. B.


\bibitem{progress}
I.~Y.~Park,
  ``Open string engineering of D-brane geometry,''
  JHEP {\bf 0808}, 026 (2008)
  [arXiv:0806.3330 [hep-th]].


\bibitem{Park:2009ki}
  I.~Y.~Park,
  ``Geometric counter-vertex for open string scattering on D-branes,''
  arXiv:0902.1279 [hep-th].





















\bibitem{gsw}
  M. B. Green, J. H. Schwarz and E. witten,
   Superstring theory, vol 1,2, Springer
\bibitem{kaku}
  M. Kaku,
Introduction to superstrings ,  Cambridge


\bibitem{Park:2008sg}
  I.~Y.~Park,
  ``One loop scattering on D-branes,''
  arXiv:0801.0218 [hep-th]v1.










\bibitem{Gonzalez-Rey:1998uh}
  F.~Gonzalez-Rey, B.~Kulik, I.~Y.~Park and M.~Rocek,
  ``Self-dual effective action of N = 4 super-Yang-Mills,''
  Nucl.\ Phys.\ B {\bf 544}, 218 (1999)
  [arXiv:hep-th/9810152].


\bibitem{Kruczenski:2006ti}
  M.~Kruczenski,
  ``Planar diagrams in light-cone gauge,''
  JHEP {\bf 0610}, 085 (2006)
  [arXiv:hep-th/0603202].
;
  M.~Kruczenski,
  ``Summing planar diagrams,''
  JHEP {\bf 0810}, 075 (2008)
  [arXiv:hep-th/0703218].










\bibitem{Das:1998ei}
  S.~R.~Das and S.~P.~Trivedi,
  ``Three brane action and the correspondence between N = 4 Yang Mills  theory
  and anti de Sitter space,''
  Phys.\ Lett.\  B {\bf 445}, 142 (1998)
  [arXiv:hep-th/9804149].
; I.~Y.~Park, A.~Sadrzadeh and T.~A.~Tran,
  ``Super Yang-Mills operators from the D3-brane action in a curved
  background,''
  Phys.\ Lett.\  B {\bf 497}, 303 (2001)
  [arXiv:hep-th/0010116].
;
  A.~J.~Nurmagambetov and I.~Y.~Park,
  ``On the M5 and the AdS(7)/CFT(6) correspondence,''
  Phys.\ Lett.\  B {\bf 524}, 185 (2002)
  [arXiv:hep-th/0110192].











\bibitem{Park:2001bm}
  I.~Y.~Park,
  ``Strong coupling limit of open strings: Born-Infeld analysis,''
  Phys.\ Rev.\  D {\bf 64}, 081901 (2001)
  [arXiv:hep-th/0106078].









\bibitem{Chepelev:1997fk}
  I.~Chepelev and A.~A.~Tseytlin,
  ``Long-distance interactions of branes: Correspondence between  supergravity
   and super Yang-Mills descriptions,''
  Nucl.\ Phys.\ B {\bf 515}, 73 (1998)
  [arXiv:hep-th/9709087].

\bibitem{Periwal:1998sn}
  V.~Periwal and R.~von Unge,
  ``Accelerating D-branes,''
  Phys.\ Lett.\  B {\bf 430}, 71 (1998)
  [arXiv:hep-th/9801121].












\bibitem{Metsaev:1998it}
  R.~R.~Metsaev and A.~A.~Tseytlin,
  ``Type IIB superstring action in AdS(5) x S(5) background,''
  Nucl.\ Phys.\  B {\bf 533}, 109 (1998)
  [arXiv:hep-th/9805028].





\bibitem{Cvetic:1999zs}
  M.~Cvetic, H.~Lu, C.~N.~Pope and K.~S.~Stelle,
  ``T-duality in the Green-Schwarz formalism, and the massless/massive IIA
  duality map,''
  Nucl.\ Phys.\  B {\bf 573}, 149 (2000)
  [arXiv:hep-th/9907202].



\bibitem{Sahakian:2004gy}
  V.~Sahakian,
  ``Closed strings in Ramond-Ramond backgrounds,''
  JHEP {\bf 0404}, 026 (2004)
  [arXiv:hep-th/0402037]
  ; 
  arXiv:hep-th/0112063.

\bibitem{Mizoguchi:2002qy}
  S.~Mizoguchi, T.~Mogami and Y.~Satoh,
  ``Penrose limits and Green-Schwarz strings,''
  Class.\ Quant.\ Grav.\  {\bf 20}, 1489 (2003)
  [arXiv:hep-th/0209043].
















\bibitem{Park:2007ev}
  I.~Y.~Park,
  ``Toward getting finite results from N=4 SYM with alpha'-corrections,''
  arXiv:0704.2853 [hep-th].












\bibitem{Gates:1986is}
  S.~J.~Gates, Jr. and S.~Vashakidze,
  ``On D = 10, N=1 supersymmetry, superspace geometry and superstring
  effects,''
  Nucl.\ Phys.\  B {\bf 291}, 172 (1987)
 ;
  E.~Bergshoeff, M.~Rakowski and E.~Sezgin,
  ``Higher Derivative Superyang-Mills Theories,''
  Phys.\ Lett.\ B {\bf 185}, 371 (1987)
 ;
  M.~Cederwall, B.~E.~W.~Nilsson and D.~Tsimpis,
  ``The structure of maximally supersymmetric Yang-Mills theory:  Constraining
  higher-order corrections,''
  JHEP {\bf 0106}, 034 (2001)
  [arXiv:hep-th/0102009]
 ;
  E.~A.~Bergshoeff, A.~Bilal, M.~de Roo and A.~Sevrin,
  ``Supersymmetric non-abelian Born-Infeld revisited,''
  JHEP {\bf 0107}, 029 (2001)
  [arXiv:hep-th/0105274]
 ;
  P.~Koerber and A.~Sevrin,
  ``The non-Abelian Born-Infeld action through order alpha'**3,''
  JHEP {\bf 0110}, 003 (2001)
  [arXiv:hep-th/0108169]





\bibitem{Fischler:1986ci}
  W.~Fischler and L.~Susskind,
  ``Dilaton Tadpoles, String Condensates And Scale Invariance,''
  Phys.\ Lett.\  B {\bf 171}, 383 (1986).


\bibitem{Das:1986dy}
  S.~R.~Das and S.~J.~Rey,
  ``Dilaton condensates and loop effects in open and closed bosonic strings,''
  Phys.\ Lett.\  B {\bf 186}, 328 (1987).

















\end{thebibliography}
\end{document}